\documentclass[12pts]{article}
\usepackage{fullpage}
\usepackage{framed,multirow}
\usepackage{graphicx}
\usepackage{lineno}
\usepackage{amsmath,amsfonts,amssymb,breqn}
\usepackage{bm}
\usepackage{graphicx}
\usepackage{subfloat}
\usepackage{subfig}
\usepackage{color}
\usepackage{hyperref}
\usepackage{setspace}

\newcommand{\vect}[1]{ \textbf{{#1}} }
\newcommand{\vers}[1]{ \hat{\bm{#1}} }

\title{Density Estimation Techniques for Multiscale Coupling of Kinetic Models of the Plasma Material Interface}

\author{Shane Keniley, Davide Curreli\footnote{SK: keniley1@illinois.edu, DC: dcurreli@illinois.edu, \emph{University of Illinois at Urbana Champaign.}}}

\date{April 7, 2018}

\begin{document}

\maketitle

\doublespacing

\begin{abstract}
In this work we analyze two classes of Density-Estimation techniques which can be used to consistently couple different kinetic models of the plasma-material interface, intended as the region of plasma immediately interacting with the first surface layers of a material wall. In particular, we handle the general problem of interfacing a continuum multi-species Vlasov-Poisson-BGK plasma model to discrete surface erosion models. The continuum model solves for the energy-angle distributions of the particles striking the surface, which are then driving the surface response. A modification to the classical Binary-Collision Approximation (BCA) method is here utilized as a prototype discrete model of the surface, to provide boundary conditions and impurity distributions representative of the material behavior during plasma irradiation. The numerical tests revealed the superior convergence properties of Kernel Density Estimation methods over Gaussian Mixture Models, with Epanechnikov-KDEs being up to two orders of magnitude faster than Gaussian-KDEs. The methodology here presented allows a self-consistent treatment of the plasma-material interface in magnetic fusion devices, including both the near-surface plasma (plasma sheath and presheath) in magnetized conditions, and surface effects such as sputtering, back-scattering, and ion implantation. The same coupling techniques can also be utilized for other discrete material models such as Molecular Dynamics. 
\end{abstract}

\section{Introduction}
\label{sec:intro}

The near-wall region of a magnetized plasma is a far-from-equilibrium system characterized by a multitude of coupled physical phenomena. Particles sputtered from the surface will interact with the plasma and either enter the plasma bulk or be redeposited, where they can both induce further sputtering events and change the constitution of the surface layers. A computational model of Plasma-Material Interactions (PMI) must be able to incorporate a kinetic description of the plasma sheath and presheath, surface response, and impurity transport in order to capture these processes. However, modeling PMI is complicated by the disparate time- and length-scales involved: relevant plasma processes occur over millimeters of length and microseconds of time, while surface interactions take place on the order of an angstrom and picosecond timescales. Accurately simulating PMI is thus not only a matter of implementing the necessary physics, but also developing the techniques for integrating physical models which necessarily operate at different scales.

Multiscale Modeling approaches this issue by using a separate model to describe each physical process, and later coupling the processes together using an opportune numerical strategy. A summary of multiscale and multiphysics research was compiled by Groen \emph{et al.} \cite{groen_survey_2012}. This paper is aimed at investigating a coupling methodology based on Density-Estimation (DE) techniques, and apply such techniques to the problem of consistently passing continuum plasma distributions to a discrete material model and back. We consider the general problem of dealing with fully-kinetic distributions for both the plasma and the material species, that is typically referred as the ``kinetic-to-kinetic'' coupling problem, as opposed to the ``kinetic-to-fluid'' coupling problem. For the plasma species we adopted a continuum multi-species electrostatic Vlasov-Poisson solver with a Bhatnagar-Gross-Krook (BGK) collision operator; such a model provides a fully-kinetic description of all relevant species in the partially-ionized conditions normally encountered close to a material surface. For the material we used a discrete surface model based on the Binary Collision Approximation (BCA) including dynamic surface composition and surface morphology, such as those implemented in the Fractal-TRIDYN code \cite{drobny_f-tridyn:_2017}. The continuum solver accurately describes the plasma sheath and presheath in front of the material surface, including full-orbit effects due to finite ion Larmor radius. The same model also provides the Ion Energy-Angle Distributions (IEAD) of the plasma particles striking the wall, which are discretized and used as an input to the surface model. The surface model in turn provides discrete energy-angle distributions of sputtered impurities, together with the distributions of reflected and implanted particles. The statistical validity of the coupling has been analyzed to ensure that the passage from discrete to continuum is performed by guaranteeing continuity of the conservation laws at the interface between the two models. 

For near-surface plasma problems involving plasma sheaths, a continuum discretization of the Vlasov-Poisson-BGK problem is generally preferable rather than discrete approaches such as Particle-in-Cells (PIC). 
PIC methods utilize a statistical sampling of the phase-space by tracking a finite number of macroparticles, \(N_p\). PIC methods have been previously adopted for near-wall simulations \cite{dejarnac_measurement_2008, birdsall_plasma_1985, khaziev_ion_2015}, but they tend to under-sample the spatial regions closest to the wall where the plasma density drops of more than an order of magnitude with respect to the upstream density. 
Furthermore, PIC codes are inherently subject to numerical noise, which can obscure features of the near-wall plasma, such as the steep gradients occurring across the magnetic presheath and inside the Debye sheath. PIC noise can be controlled in a number of ways, such as increasing the number of computational particles \(N_p\), but it typically diminishes only by a factor of \(\sqrt{N_p}\). 
In contrast, continuum kinetic solvers directly discretize the Vlasov-Poisson-BGK problem in phase-space with a mesh. They are more computationally expensive than equivalent PIC models, but avoid the numerical noise problem of PIC codes. The method allows higher resolution even in regions of large gradients such as the plasma sheath. Eulerian schemes applied to the problem have been studied with semi-Lagrangian methods \cite{parker_convected_1997,sonnendrucker_semi-lagrangian_1999,cheng_integration_1976,qiu_conservative_2010}, discontinuous Galerkin schemes \cite{cagas_continuum_2017}, and finite difference methods \cite{filbet_mixed_2014}. In this work we will adopt a high-order Cheng-Knorr Strang splitting of the Vlasov operator. 
The erosion process itself is traditionally modeled using either Molecular Dynamics (MD) or a BCA approach. The advantage of BCA over MD is mainly computational, since the ion-material interaction is reduced from a multi-body problem to a Monte Carlo sequence of binary collisions, allowing faster computations and straightforward parallelization. Furthermore, the sputtering yields computed via BCA have been shown to have good agreement with experimental measurements even at low-energy in regimes close to the sputtering threshold \cite{yamamura_low-energy_1985}. For a review on atomic collisions in amorphous targets with BCA methods, see \cite{DROBNY2018301}.  

Thanks to the coupling methodology developed in this paper, a self-consistent kinetic model of the near-surface plasma (plasma sheath and presheath) including material response has been obtained. The model allows kinetic analyses of the near-surface plasma behavior, and numerical predictions of the erosion, reflection, and implantation rates during PMI activity. In the following sections, first we present the model equations of both the plasma and the material in order to establish the necessary exchange of information between the two regions, and then we describe the method of dynamically coupling the multi-species Boltzmann-Poisson plasma treated with finite differences to the BCA model describing the plasma-facing surface (Section \ref{sec:equations}). Tests to ensure the statistical validity of the coupling method are presented in section \ref{sec:tests}, including an example of the dynamically-coupled model of a plasma sheath.

\section{Model Equations}
\label{sec:equations}

\subsection{Boltzmann-Poisson problem of a magnetized plasma}
\label{sec:boltpoiss}

Consider a volume $dV = d \vect{x} d \vect{v}$ of the phase space $(\vect{x},\vect{v})$, where $\vect{x} = (x_1,x_2,x_3)$ is the position vector and $\vect{v} = (v_1,v_2,v_3)$ is the velocity vector, populated with $N_s$ plasma particles of species $s$ in which the density of particles is sufficiently large that in an infinitesimally small region the condition $N_s d \vect{x} d \vect{v} \gg 1$ is satisfied. In this region, the average distribution of particles in space and velocity at time $t$ may be described by the distribution function $f_s(\vect{x},\vect{v},t)$. The time evolution of such a distribution function over the phase space $(\vect{x},\vect{v})$ is governed by the Boltzmann-Poisson integro-differential problem \cite{boltzmann_further_2003}: 
\begin{align}
& \frac{\partial f_s}{\partial t} 
+ \vect{v} \cdot \nabla_{\vect{x}} f_s 
+ \frac{q_s}{m_s} ( -\nabla_{\vect{x}} \phi +  \vect{v}\times\vect{B}) \cdot \nabla_{\vect{v}} f_s 
= \left( \frac{\partial f_s}{\partial t} \right)_{coll} \label{eq:bke1}\\
& \epsilon_0 \nabla^2_{\vect{x}} \phi = q_e n_0 \exp \left( \frac{q_e \phi}{KT_e} \right) -\sum_s q_s \int f_s d \vect{v} \label{eq:bke2}
\end{align}
where $m_s$ and $q_s$ are the mass and the electric charge of all massive species $s$ (ions and neutrals), $\phi$ the electric potential, and $\vect{B}$ the magnetic field.
The collision operator at the right-hand side of Eq.~\ref{eq:bke1} accounts for the Fokker-Planck and atomic modifications to the distribution function $f_s$ due to collisional processes. In first approximation the collision operator within the plasma region can be treated as a Bhatnagar-Gross-Krook (BGK) operator \cite{bhatnagar_model_1954}, 
\begin{equation}
\left(\frac{\partial f_s}{\partial t}\right)_{coll} = \nu_{bgk} (f_{s,0}-f_s(\vect{x},\vect{v},t))
\label{EqBGK1} 
\end{equation}
allowing explicit handling of elastic scattering, electron-neutral collisions, and charge-exchange between ions and neutrals, where each process is described by its relaxation frequency $\nu_{bgk}$. Collisions in the material region are treated with a dedicated binary-collision operator accounting for ion-matter interaction, described in Section~\ref{sec:bca}.

The Vlasov-Poisson-BGK problem of Eqs.~\eqref{eq:bke1}-\eqref{eq:bke2}-\eqref{EqBGK1} is numerically solved in one spatial dimension and three velocity dimensions (1D3V) by means of a finite-volume discretization of the phase-space with a Cheng-Knorr \cite{cheng_integration_1976,marchuck_splitting_2015} high-order Strang splitting of the Vlasov operator \cite{strang_construction_1968,einkemmer_convergence_2014}:
\begin{enumerate}
	\item $\partial_t f_s + v_x \partial_x f_s = 0$ \hspace{8.8em} x-advection $(\Delta t / 2)$
	\item $\nabla^2 \phi = -\rho / \epsilon_0$  \hspace{10.5em} Poisson solver
	\item $\partial_t f_s + \tfrac{q_s}{m_s}(E_x + v_y B_z - v_z B_y)\partial_{v_x} f_s = 0$ \hspace{1em} $v_x$-advection $(\Delta t)$
	\item $\partial_t f_s + \tfrac{q_s}{m_s}(v_z B_x - v_x B_z)\partial_{v_y} f_s = 0$ \hspace{3em} $v_y$-advection $(\Delta t)$
	\item $\partial_t f_s + \tfrac{q_s}{m_s}(v_x B_y - v_y B_x)\partial_{v_z} f_s = 0$ \hspace{3em} $v_z$-advection $(\Delta t)$
	\item $\partial_t f_s + v_x \partial_x f_s = 0$ \hspace{8.8em} x-advection $(\Delta t / 2)$
	\item $\partial_t f_s = \nu_{bgk} (f_0 - f(x, v))$ \hspace{6.25em} Collision operator
\end{enumerate}
Operator splitting reduces the Vlasov-Poisson-BGK problem of Eqs.~\eqref{eq:bke1}-\eqref{eq:bke2}-\eqref{EqBGK1} from a multidimensional integro-differential problem to a sequence of problems of lower dimensionality, each one with uniquely-defined nature (elliptic, hyperbolic, quadrature). 
High-order upwind schemes are used for the hyperbolic sections of the problem, finite-difference discretization for the elliptic portion, and high-order Simpson schemes for the quadratures.

A key feature desired for plasma sheath simulations is that the plasma always flows from the bulk toward the boundary. Upwind-biased numerical schemes exploit this directionality, achieving the highest accuracy per gridpoint in a finite difference stencil \cite{van_leer_upwind_2006}. First order upwind-biased methods are stable and convergent, but are too dissipative without a highly-refined grid. This work will instead focus on high-order upwind methods. In particular, since the rate at which error is decreased with increasing resolution is faster by an order of magnitude for a high-order method, the fourth-order upwind scheme will be used for our simulations. The third-order method has a less strict stability condition which would allow for larger timesteps to be used \cite{zha_accuracy_2003}, but the error control of the higher order scheme was considered more beneficial than including a larger timestep. High order time approximations are also necessary for the time-pushing in order to achieve the same accuracy of the grid upwinding, the most common example of which are the set of Runge-Kutta algorithms \cite{jameson_numerical_1981}. The present work utilizes a fourth-order explicit Runge-Kutta (RK4) method. 

The simulation of a near-wall plasma requires a grid size small enough to resolve the plasma sheath, which requires a resolution on the order of the Debye length ($\Delta x = O(\lambda_{De})$ ). However, since using such a high resolution across the entire domain is computationally expensive, a nonuniform grid that is highly refined in only a small region is desirable. In this work we adopt a hyperbolic-tangent mesh refinement suitable to the plasma sheath problem, providing fine resolution on one end of the domain for the plasma sheath and coarse resolution on the other for the plasma bulk, with a gradual transition between the two regions \cite{devaux_vlasov_2006}:
\begin{equation}
g(s) = \Delta x_1 + \frac{\Delta x_2 - \Delta x_1}{2} (1 + \tanh{[\alpha(s-s0)]})
\label{EqNonuniform}
\end{equation}
where $\Delta x_1$ is the refined grid step near the wall, $\Delta x_2$ is the grid step in the plasma bulk, $\alpha$ is the steepness of the transition region, and $s_0$ is the location of the transition. The adoption of a non-uniform spatial grid, Eq.~\ref{EqNonuniform} requires a modification of the finite difference schemes used for the discretization of the operators. The approximate (discrete) solution $u_i = u_i(x)$ of a generic function of the spatial coordinate $\hat{u} = \hat{u}(x)$ may be defined on a general numerical grid with nonuniform grid spacing with an $N_g$-order polynomial, where $N_g$ is the order of accuracy of the resulting numerical scheme. As an example, a second-order upwind-biased discretization at a point $i$ is represented by the following set of polynomials: 
\begin{align}
\hat{u} = ax^2 + bx + c \\
u_{i} = c \\
u_{i-1} = a (-h_{i-1})^2 + b (-h_{i-1}) + c \\
u_{i-2} = a (-2 h_{i-2})^2 + b (-2 h_{i-2}) + c
\end{align}
with undefined coefficients $a$, $b,$ and $c$. The polynomial fit defines a linear system of equations. An expression for the coefficients $a$, $b$, and $c$ may be found by inverting the resulting matrix, and the numerical discretization at any point is then derived by taking the derivative of $\hat{u}$: 
\begin{align*}
\frac{d \hat{u}}{d x} &= 2ax + b \\
\frac{d^2 \hat{u}}{d x^2} &= 2a 
\end{align*}
Thus the second-order upwind-biased discretization of the first derivative is defined as the coefficient $b$. For a uniform grid spacing, it is simple to verify that the parametric fit method reproduces the traditional second-order upwind scheme. The same treatment is used to derive nonuniform spacing for the third- and fourth-order upwind schemes, along with the second- and fourth-order central difference schemes for Eq. \ref{eq:bke2}.

\subsection{Surface Erosion and Ion Implantation Model}
\label{sec:bca}

The trajectory of each individual ion leaving the plasma and entering into a material surface of particle density $N$ follows a random walk characterized by: (1) large deflections of the incoming particle at each encounter with the nuclei, and (2) continuous energy losses between one nuclear encounter and the following due to inelastic electronic interactions. 
Here we consider the problem of evaluating the change to the distributions $f_s$ needed for surface erosion and ion implantation considering the following Fokker-Planck problem:
\begin{equation}
\nabla \cdot \left[ \mathcal{V} f_s - \nabla ( \mathcal{D} f_s) \right] = S
\label{EqFokkerPlanck}
\end{equation}
The problem of Eq.~\ref{EqFokkerPlanck} describes the drift and diffusion of the plasma and material particles traveling through the material under a source $S$ determined by the plasma sheath. Eq.~\ref{EqFokkerPlanck} can be recast \cite{kloeden_numerical_1992} as a Stochastic Differential Equation (SDE), and resolved with a Lagrangian approach by reconstructing the distributions from a large number of uncorrelated particle histories experiencing a Wiener process. It can be shown that the Fokker-Planck problem of Eq.~\ref{EqFokkerPlanck} is equivalent to the following Ito SDE:
\begin{equation}
d \vect{x}(t) = \mathcal{V}(\vect{x},t) \: dt + \mathcal{D}(\vect{x},t) \: \vect{dW}(t)
\label{EqIto1}
\end{equation}
where $\vect{x}$ is the the particle position vector, $\mathcal{V}(\vect{x},t)$ the drift vector, $\mathcal{D}(\vect{x},t)$ the diffusion tensor, and $\vect{dW}(t)$ is the differential element of a multi-dimensional Wiener process. 
When Eq.~\ref{EqIto1} is projected on a reference frame $(\vers{e}_1,\vers{e}_2,\vers{e}_3)$ having the third axis $\vers{e}_1$ parallel to the surface normal, $\vers{e}_1 // \vers{n}$, the following Ito equation is obtained (written by components):
\begin{equation}
d \vect{x}(t) = v_i \: \vers{e}_i \: dt + \sqrt{ 2 D_i } \vers{e}_i \: d W_i, \quad \quad i=1,..,3
\label{EqIto2}
\end{equation}
where the Einstein notation on repeated indexes is adopted; the components of the generalized drift are $\mathcal{V} = (v_1,v_2,v_3)$; and the three orthonormal components of the diffusion tensor are $D_i = eig ( \frac{1}{2} \mathcal{D} \mathcal{D}^{T} ) = (D_1,D_2,D_3)$. A discrete first-order approximation to the Ito problem of Eq.~\ref{EqIto2} is then readily obtained from the Euler-Maruyama scheme through the Markov chain $\vect{x}^k = \vect{x} ( t^k)$ at discrete time intervals $t^k = k \: \Delta t$:
\begin{equation}
\vect{x}^{k+1} = \vect{x}^{k} + v_i^k \: \hat{\vect{e}}^{k} \Delta t + \sqrt{ 2 D_i } \vers{e}_i \: \Delta W_i, \quad i = 1,..,3
\label{EqIto3}
\end{equation}
where the random variables $\Delta W_i$ $(i=1,..,3)$
\begin{equation}
\Delta W_i = W_i(t^k + \Delta t) - W_i (t^k)
\end{equation}
are either uniformly distributed random variables on a circle (for angular coordinates), or normally distributed random variables (for translations) with zero expected value and variance $\Delta t$. 

In Binary Collision Approximation codes such as TRIM \cite{Ziegler198595}, TRIDYN \cite{MOLLER1984814}, SDtrimSP \cite{Mutzke09sdtrimsp-2d:simulation}, and Fractal-TRIDYN \cite{drobny_f-tridyn:_2017}, the following simplifying assumptions are made to Eq.~\ref{EqIto3}, 
\begin{align}
& v_i^k \: \Delta t = \lambda - s \quad (i=1,2,3), \label{EqBCA1}\\
& \epsilon^{k+1} = \epsilon^k - \Delta \epsilon_{loss}^k, \label{EqBCA3}
\end{align}
In Eq.~\ref{EqBCA1} the total path covered by the particle between two subsequent nuclear encounters is given by the difference of the local mean free path $\lambda = N^{-1/3}$ and the distance of asymptotic deflection, $s = R \sin(\theta/2)$, where $R$ is the distance of closest approach and $\theta$ is the deflection angle of the ion trajectory expressed in the Center-of-Mass (CoM) reference frame. 
The distance of closest approach $R$ depends on the details of the ion-nucleus interaction potential $V$, which can be written as the product of a Coulomb-like potential of constant $V_o = q_e^2 Z_1 Z_2 / (4 \pi \epsilon_0)$, and a nuclear screening function $\phi(\xi)$ expressed as a series of Yukawa-like exponentials (typically 3-4 terms), 
\begin{equation}
V = \frac{V_o}{R} \phi(\xi), \quad \phi(\xi) = \sum_i c_i e^{d_i \xi}, \quad \text{where} \quad \xi = R/a.  \label{EqBCA5}
\end{equation}
The numerical value of $R$ is then obtained by means of a Newton-Raphson scheme from the solution of the following non-linear scalar equation (obtained from conservation of energy and angular momentum),
\begin{equation}
\epsilon_r p^2 + a R \phi(\xi) - a R \epsilon_r \xi = 0,  \label{EqBCA6} 
\end{equation}
where $p^2 = r_o/(\lambda \pi N)$ is the collisional radius (or impact parameter), $r_o \in (0,1]$ a uniformly-distributed random number, $\epsilon_r = (a / V_o)\epsilon_{CoM} $ the reduced energy, $\epsilon_{CoM} = \epsilon m_2/(m_1 + m_2)$ the CoM energy, $a = a_0 (9\pi^2/128)^{1/3} (Z_1^{1/2} + Z_2^{1/2})^{-2/3}$ the atomic screening distance, and $a_0$ the Bohr radius, $a_0 = \hbar / m_e c \alpha$. The deflection angle $\theta$ is calculated in the CoM reference frame by solving the geometry of the binary collision event between the ion and the nucleus; the value of $\theta$ in the laboratory frame is then given by:
\begin{equation}
\tan \Psi = \frac{\sin \theta}{m_2/m_1 + \cos \theta} \label{EqBCA7} 
\end{equation}
The diffusion tensor $\mathcal{D} = \mathcal{D} (\hat{\vect{u}}^{k} | \Psi, \Phi )$ is obtained from the collision operator of binary collisions having interaction potential as in Eq.~\ref{EqBCA5}, where $\hat{\vect{u}}^{k} = (\cos \alpha, \cos \beta, \cos \gamma)$ is the vector of the direction angles at the current iteration $k$, the angle $\Psi$ is the the deflection angle in the laboratory frame (Eq.~\ref{EqBCA7}), and $\Phi = 2 \pi r_o$ (azimuthal deflection angle) is a uniformly distributed  random variable on a circle. 
Additional diffusion processes, such as temperature effects, are typically neglected in the BCA approach. Including them in the scheme of Eq.~\ref{EqIto3} is a straightforward extension of the current treatment, and will not be considered in the present work. 

The integral energy loss $\epsilon_{loss}$ between one collision and the following (Eq.~\ref{EqBCA3}) is given by the sum of the energy transferred to the nucleus $\epsilon_{n}$ and the inelastic electronic losses (both local $\epsilon_{loc}$ and non-local $\epsilon_{nl}$), 
\begin{equation}
\Delta \epsilon_{loss} = \epsilon_{n} + \epsilon_{loc} + \epsilon_{nl} 
\end{equation}
where the nuclear energy transfer is provided by a simple balance of the energy-momentum of the two colliding particles, $\epsilon_{n} = 4 \epsilon m_1 m_2 / (m_1 + m_2)^2 \sin^2 (\theta/2)$, and the electronic losses $\epsilon_{nl} = (\lambda - s) N S_{el}$ and $\epsilon_{loc} = d_1^2 \exp(d_1 \xi) \: S_{el}/(2 \pi a^2)$ are both functions of Lindhard's \cite{lindhard1963} semi-classical electronic stopping power $S_{el} \propto \sqrt{\epsilon}$. 

\begin{figure}[htp]
\centering
	\subfloat[Atomic Sputtering] {
		\includegraphics[width=0.3\textwidth]{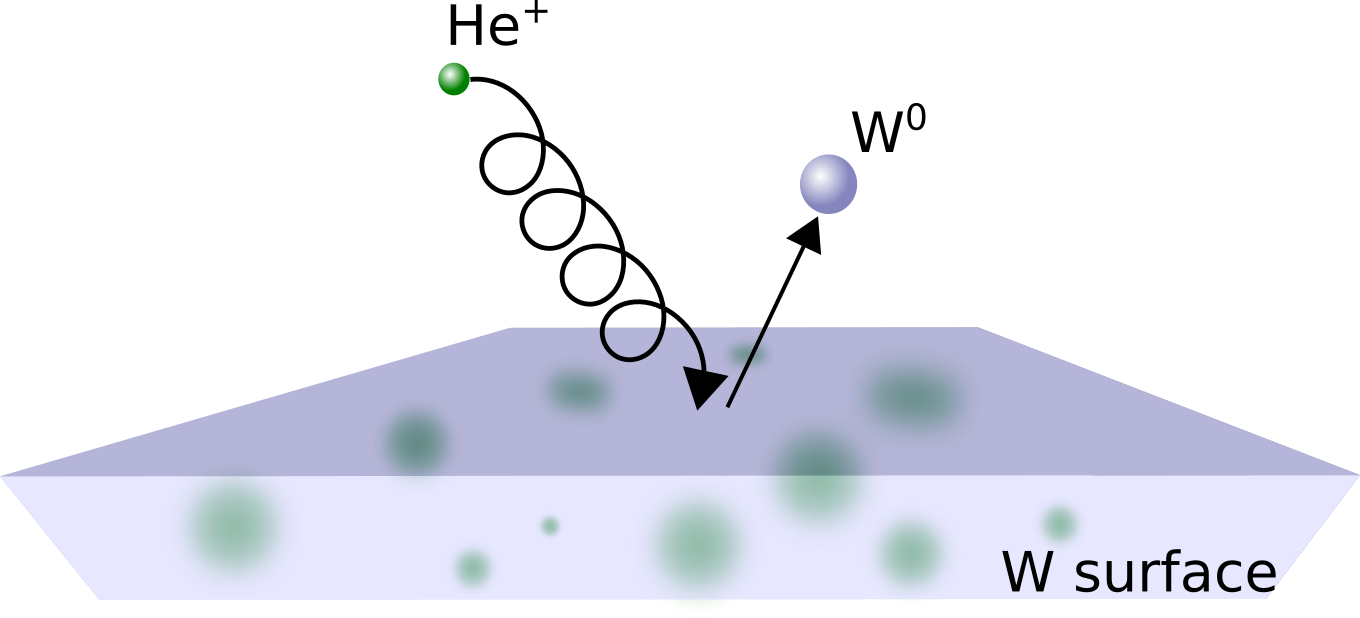}
		\label{fig:pmi_sputtering}
	}
	\subfloat[Self Sputtering] {
		\includegraphics[width=0.3\textwidth]{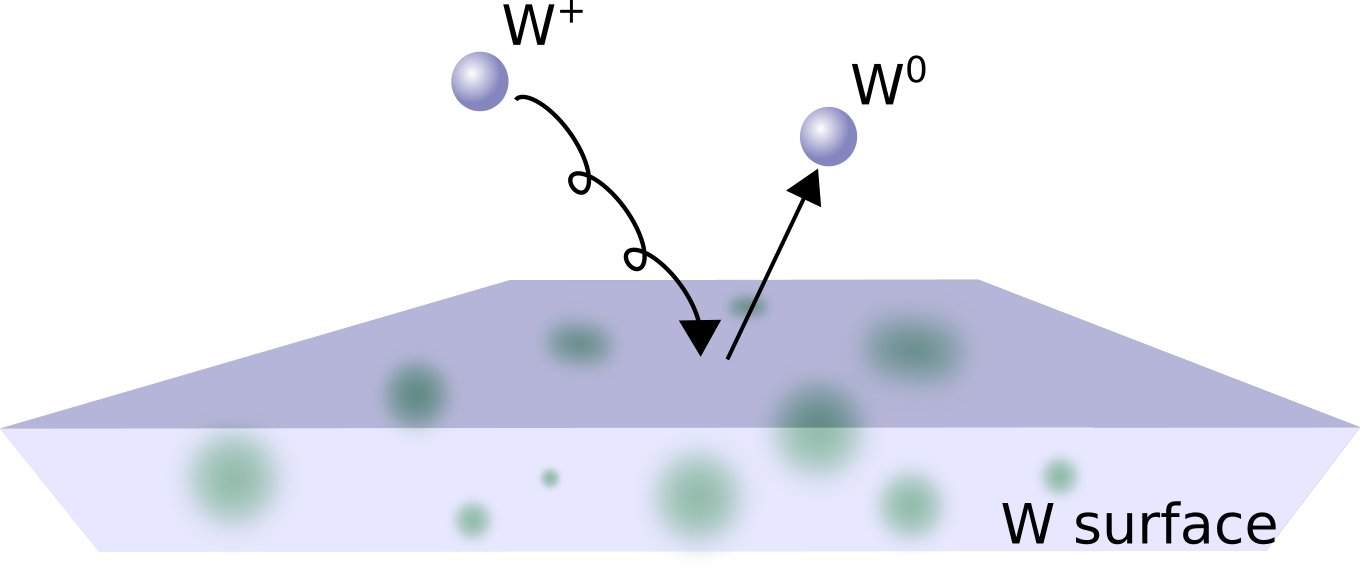}
		\label{fig:pmi_self}
	}
	\subfloat[Gas Sputtering] {
		\includegraphics[width=0.3\textwidth]{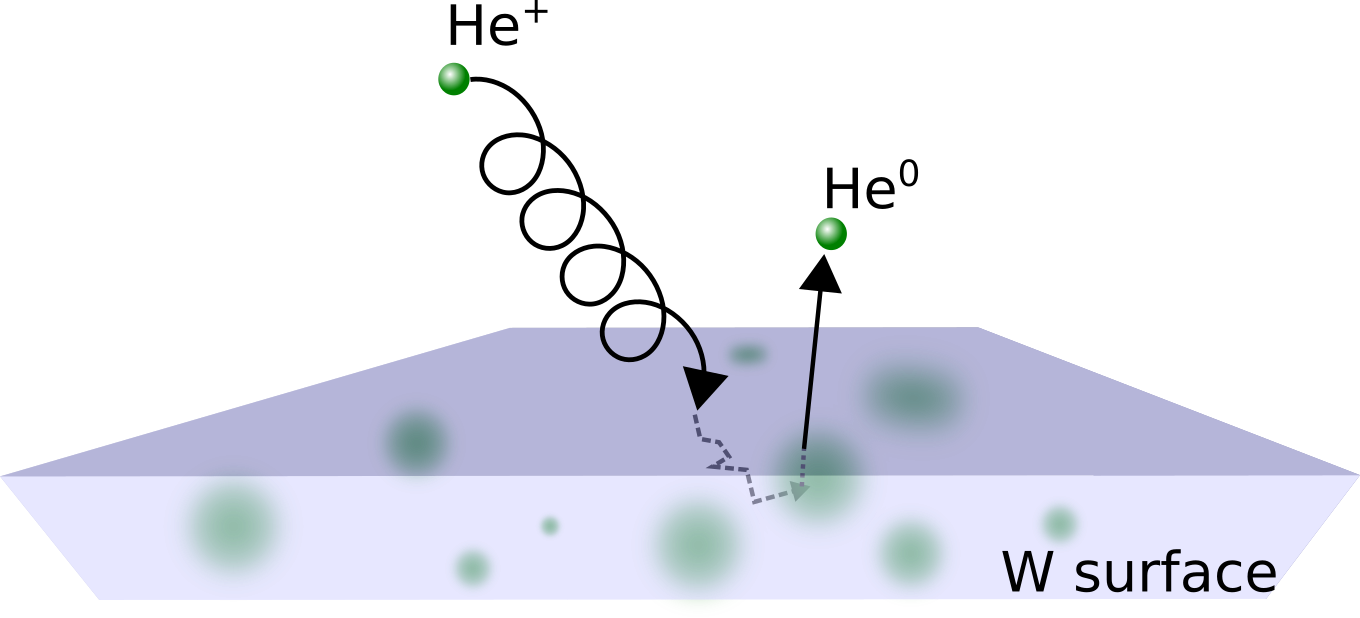}
		\label{fig:pmi_gas}
	}
    
	\subfloat[Reflection] {
		\includegraphics[width=0.3\textwidth]{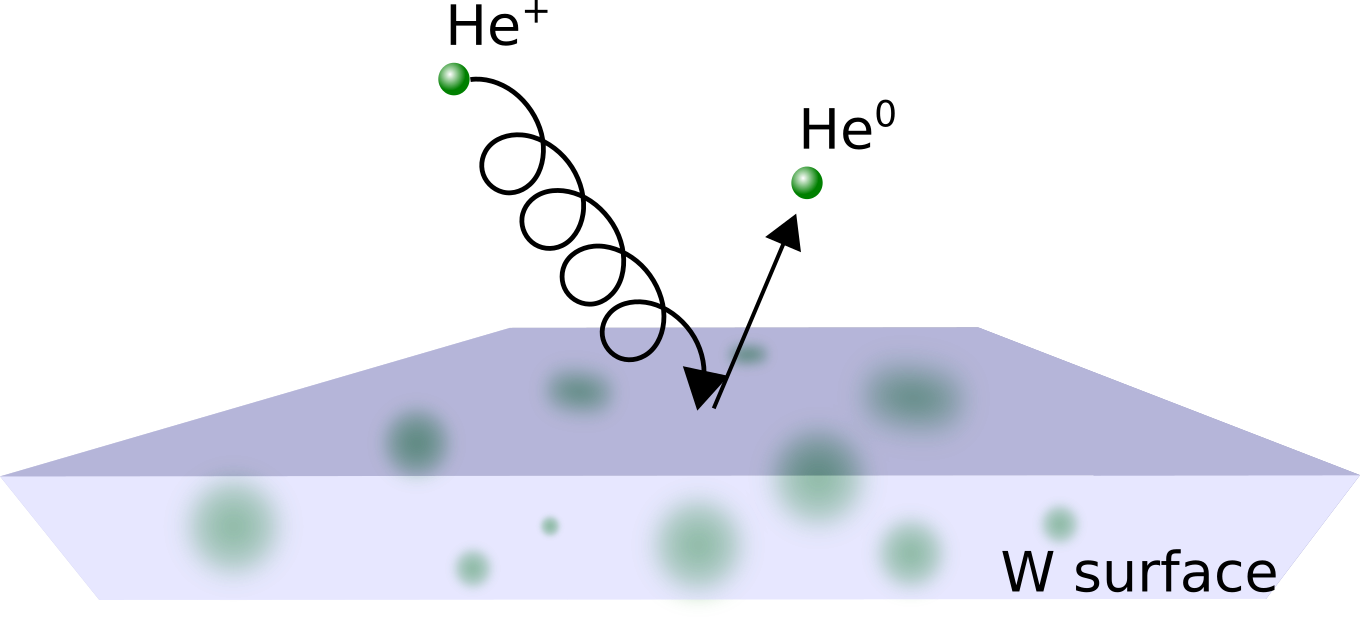}
		\label{fig:pmi_reflection}
	}
	\subfloat[Implantation] {
		\includegraphics[width=0.3\textwidth]{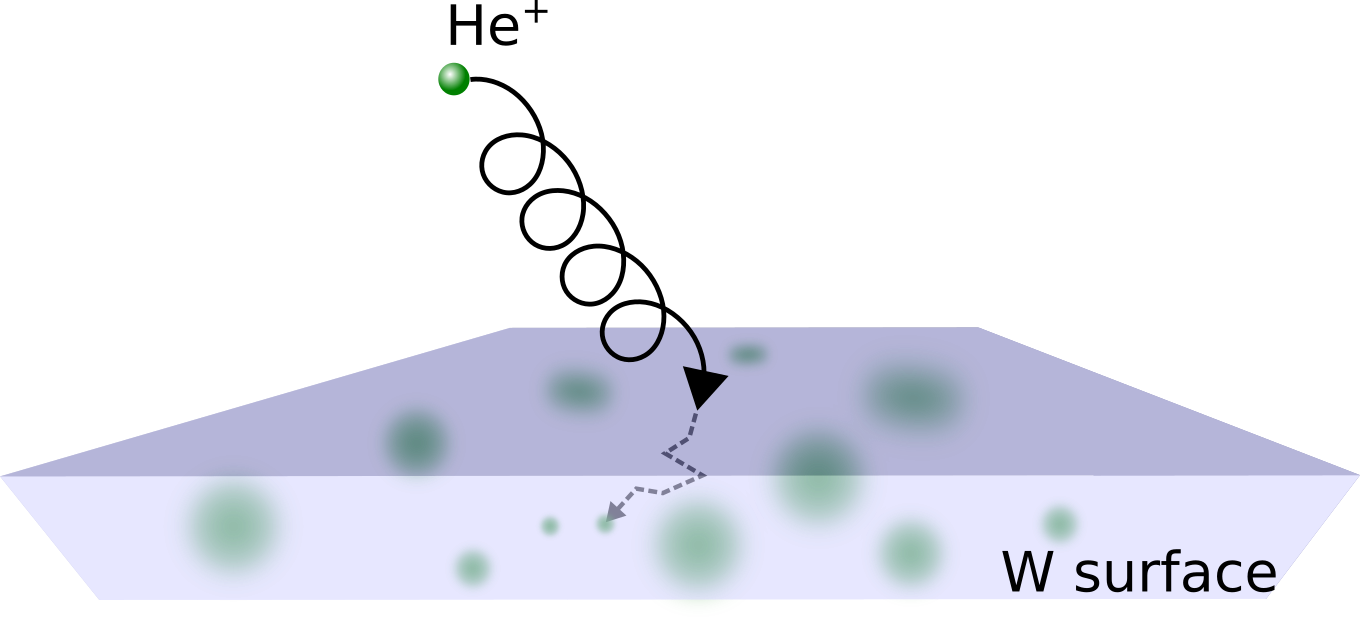}
		\label{fig:pmi_implantation}
	}
	\subfloat[Prompt Reemission] {
		\includegraphics[width=0.3\textwidth]{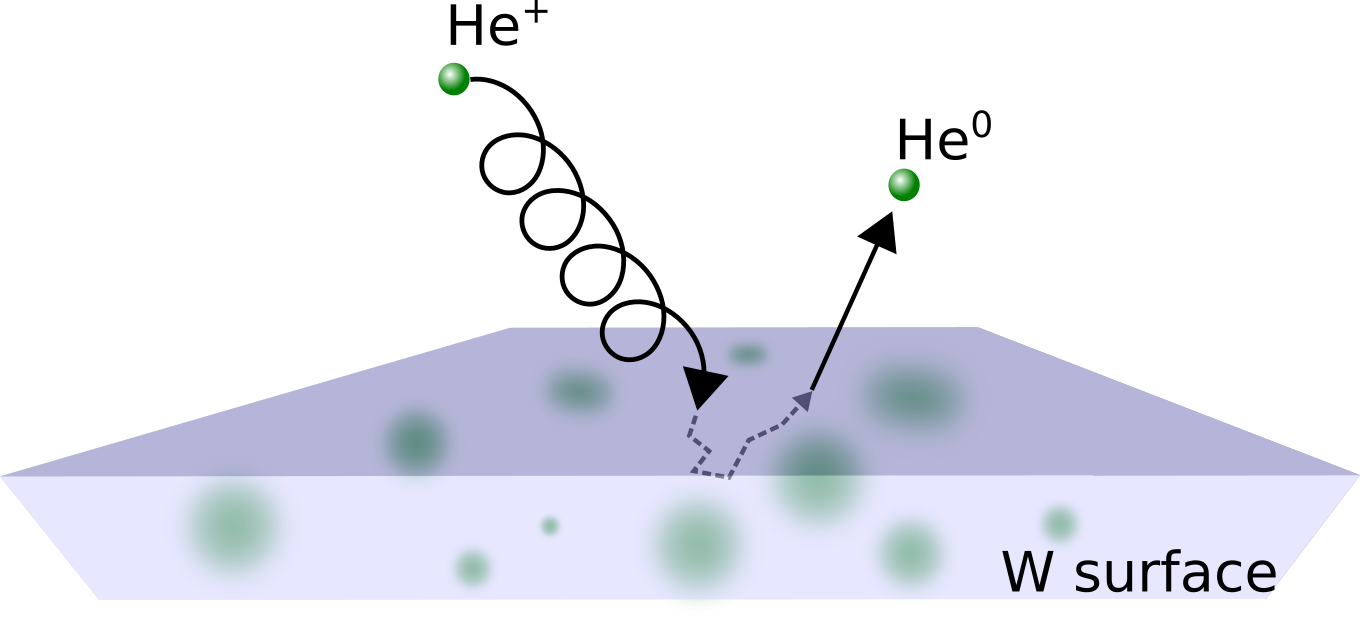}
		\label{fig:pmi_prompt}
	}
	
	\caption{Illustrations of the plasma-surface interaction processes  modeled by the method described in Sec.~\ref{sec:bca}. \label{fig:pmi}}

\end{figure}

The distributions of interest $f_s$ can be reconstructed from a large number of Markov chains ($N_{mc}$), typically $N_{mc} \sim \mathcal{O}(10^4-10^5)$, such as those in Eq.~\ref{EqIto3} with discrete displacements described by Eqs.~\ref{EqBCA1}--\ref{EqBCA3}. It is relevant to highlight that the individual Markov chains retain a purely Lagrangian character. A computational mesh is only required to collect the cumulative contribution of each individual trajectory and reconstruct the distributions $f_s$. This is traditionally accomplished by means of a track length estimator. For a mesh made of $c=1,...,N_c$ cells, an estimate of the  distribution $f_s$ within each cell is obtained from the cumulative of the time $\tau_{jc}$ spent by a particle $j$ visiting the cell $c$:
\begin{equation}
f_{s,c} = \frac{1}{V_c} \sum _{j=1} ^{N_{mc}} w_j \tau_{jc}, \quad (c = 1, ..., N_c)
\end{equation}
where $V_c$ is the volume of the $c$-esim cell and $w_j$ the weight of the particle. The method just described in this section offers a linearized solution to the Ito problem of Eq.~\ref{EqIto2}, and enables the treatment of the following plasma-surface interaction processes (schematically depicted in Figure \ref{fig:pmi} for the case of a helium plasma impacting on an inhomogeneous tungsten wall): (1) material sputtering, (2) self-sputtering, (3) sputtering of implanted gas, (4) reflection, (5) implantation, and (6) prompt reemission. 

\subsection{Plasma-Wall Coupling}

In this section we describe a general method to couple the continuum plasma model described in Sec.~\ref{sec:boltpoiss} to the discrete surface model reported in Sec.~\ref{sec:bca}. The possibility to dynamically couple the two models and simulate the plasma-wall system ``on-the-fly'' strictly depends on the efficiency of the numerical scheme adopted for coupling. In this section we discuss the model equations used to merge the two physical models, with particular care to numerical efficiency. The coupling methodology requires: (1) a geometrical definition of the surface shape; (2) a consistent strategy to convert particle distributions to continuum distributions; (3) consistency criteria to be satisfied in order to enforce the conservation laws at the interface. 

In the most general case, the shape of a material surface is a three-dimensional surface of topological dimension $2 \leq D < 3$, where $D=2$ corresponds to an ideal flat surface, and intermediate dimensions between two and three are for rough surfaces of fractal dimension $D$ [\textbf{cite}]. The geometrical location of a surface is defined by an implicit function of space $G(\textbf{x}) = 0$. The simplest case of a flat surface corresponds to the function $x-L=0$, with $L$ the location of the wall and $x$ a scalar coordinate perpendicular to the wall. More complex surfaces (e.g. fractals) can be defined using the same strategy as well \cite{drobny_f-tridyn:_2017}. 
The function $G$ also provides a logical condition of surface crossing, useful to determine whether a particle or a plasma fluid element is inside or outside the wall,
\begin{align}
& G(\textbf{x}) > 0 \quad : \quad \text{Outside Wall} \label{Eq1LogicWall} \\
& G(\textbf{x}) < 0 \quad : \quad \text{Inside Wall} \label{Eq2LogicWall} 
\end{align}
The logical condition of surface crossing (Eqs.~\ref{Eq1LogicWall}-\ref{Eq2LogicWall}) can be applied to each of the $N_{mc}$ individual Markov chains of Eq.~\ref{EqIto3}, $G(\vect{x}^k) > 0$, in order to determine whether a particle traveling inside the wall has crossed the surface and is thus released into the plasma. This logical test provides a number $j=1,..,N_{ex}$ of particles released by the surface, where in general $N_{ex} \leq N_{mc}$. The state vector of the particles leaving the surface at time $t^k$ is $\vect{q}_j (t^k) = \left( \vect{x}_j(t^k), \vect{v}_j(t^k) \right)$, where the particle velocity at time $t^k$ is given by $\vect{v}_j = \vers{u}_j ( 2 \epsilon_j / M_j ) ^{1/2}$. The fractions of sputtered particles $Y_s$, backscattered particles $B_s$ and  implanted particles $I_s$ are given respectively by (per each species $s$):
\begin{equation}
Y_s = \sum_{p} \frac{N_{ex,p}}{N_{mc}}, \quad B_s = \frac{N_{ex,s}}{N_{mc}}, \quad I_s = \frac{N_{mc} - N_{ex,s}}{N_{mc}}
\end{equation}
where $p$ is an index running over all secondary particles produced by the species $s$. The particle ensemble $\vect{q}_j$ $(j=1,..,N_{ex})$ is converted to a continuum distribution by means of the Density Estimation operator $\mathcal{K} = \mathcal{K}(\vect{q}_j)$, which provides an estimate $\hat{f}$ of the continuum distribution function $f$ using information from discrete particle vectors. For particles leaving the surface at time $t^k$ at those locations $\vect{x}$ specified by $G(\vect{x}) = 0$, the reconstructed velocity distribution at the interface is provided by an operator $\mathcal{K} = \mathcal{K}(\vect{v}_j)$ which is a function of the particle velocity only,
\begin{equation}
\hat{f}_s ( \vect{x}, \vect{v}, t^k ) = \frac{1}{N_{ex} \Delta v} \sum_j^{N_{ex}} \mathcal{K} \left( \frac{\vect{v}- \vect{v}_j}{\Delta v} \right), \quad \text{at } \: \: G(\vect{x}) = 0
\end{equation}
where $\Delta v$ is the window width (bandwidth) along the velocity coordinates, and $\vect{v}$ is a continuum velocity coordinate. The choice of the density estimators $\mathcal{K}$ and of the optimal bandwidth $\Delta v$ will be discussed in detail in Sec.~\ref{sec:densest}. Note that the reconstructed distribution $\hat{f}_s$ is by definition normalized to one, $\int \hat{f}_s d \vect{v}= 1$, so that in order to restore physical units it must be multiplied by the density of particles backscattered by the surface $G(\vect{x}) = 0$, 
\begin{equation}
\Delta f_s ( \vect{x}, \vect{v}, t^k ) = \Delta n_{s} \hat{f}_s ( \vect{x}, \vect{v}, t^k ), \quad \text{for } \:\: G(\vect{x}) = 0, \:\: \vect{v} \cdot \vers{n} > 0
\end{equation}
where $\vers{n}$ is the surface normal, and the density $\Delta n_{s}$ is given by
\begin{equation}
\Delta n_{s} (\vect{x})=  \sum _{s'} B_{s'} \int \Delta f_{s'} ( \vect{x}, \vect{v}, t^k ) d \vect{v}, \quad \text{for } \:\: G(\vect{x}) = 0, \:\: \vect{v} \cdot \vers{n} < 0
\end{equation}
where $s'$ is an index that runs over all species contributing to backscattering of species $s$ (sputtering plus reflection), including self-interactions of species $s$ with itself. Here the quantity $\Delta f_{s'}$ denotes the fraction of particles entering the wall between time $t^k$ and time $t^{k+1} = t^k + \Delta t$,
\begin{equation}
\Delta f_{s'} = \int _{t^k} ^{t^{k+1}} f_{s'}( \vect{x}, \vect{v}, t ) dt, \quad \text{for } \:\: G(\vect{x}) = 0, \:\: \vect{v} \cdot \vers{n} < 0
\end{equation}
Note the opposite sign of the projection $\vect{v} \cdot \vers{n}$, which is positive for particles leaving the wall, and negative for particles entering the wall on the surface $G(\vect{x}) = 0$, obeying the convention 
\begin{align}
&\vect{v} \cdot \vers{n} > 0 \quad : \quad \text{Particles leaving the wall} \\ 
&\vect{v} \cdot \vers{n} < 0 \quad : \quad \text{Particles entering the wall}
\end{align}

\subsection{Continuum-to-Particle: Rejection Sampling}

The conversion from a continuous distribution $f_s(\vect{x}, \vect{v}, t)$ to a particle distribution $\vect{q}_j$ $(j=1,..,N_{mc})$ was obtained through a simple rejection sampling technique \cite{Robert2004}. Tests of best utilization of different instrument distributions (uniform, normal) suggested that a multivariate normal distribution can achieve maximum utilization in the range of 60-70\% by means of a covariance scaling factor comprised between 1 and 2.

\subsection{Particle-to-Continuum: Probability Density Estimators}
\label{sec:densest}

We have considered two classes of Density Estimation techniques for the calculation of the distribution $\hat{f}$ from the particle ensemble $\vect{q}_j$: (1) the parametric Gaussian Mixture Model (GMM), and (2) the nonparametric Kernel Density Estimation (KDE). The two classes are representative of the two modern approaches to density estimation, which may be split into: \emph{parametric}, which assumes the Probability Density Function (PDF) may be fitted by some parametric family of functions, and \emph{non-parametric}, which can be applied to an arbitrary data set without having any prior knowledge of the shape or the behavior of the underlying distribution function. Here we briefly describe the two techniques and the selection criteria for optimal density estimate. 

The Gaussian Mixture Model is parametrized by the component weights $w_g$, the component means $\mu_g$, and the variance $\sigma_g$, where the index $g$ runs over the number of Gaussian functions  $g=1,...,N_{gmm}$. The component weights are constructed such that $\sum_{g=1}^{N_{gmm}} w_g = 1$. For example in one dimension the expression of $\hat{f}$ is given by
\begin{equation}
\hat{f}(x) = \sum _{g=1} ^{N_{gmm}} w_j \mathcal{N}( x | \mu_g, \sigma_g), \quad \mathcal{N}( x | \mu_g, \sigma_g) = \frac{1}{\sqrt{2 \pi}} \frac{1}{\sigma_g} \exp \left( -\frac{(x - \mu_g)^2}{2 \sigma_g^2} \right) \quad \quad \text{(1D)} \label{eq:gmm}
\end{equation}
The Gaussian mixture model has been paired with an expectation maximization algorithm \cite{dempster_maximum_1977,li_mixture_2000} to estimate the values of $\mu_g$ and $\sigma_g$ maximizing the probability of the particle data $\vect{q}_j$. The optimal number of Gaussian $N_{gmm}$ was decided using the Akaike information criterion \cite{Akaike1974}.  

In Kernel Density Estimation (KDE), the shape of the kernel is variable, but the most widely used kernels are PDFs themselves and thus have the desirable property that their integral is equal to one, $\int_{-\infty}^{\infty} \mathcal{K}(y) dy = 1$. With a given kernel $\mathcal{K}(y)$, the distribution $\hat{f}$ can be constructed as \cite{silverman_density_1986}
\begin{equation}
\hat{f}(x) = \frac{1}{N_{ex} \Delta x} \sum_{j=1}^{N_{ex}} \mathcal{K}\left( \frac{x-x_j}{\Delta x} \right),
\label{eq_kde}
\end{equation}
where $\Delta x$ is the bandwidth. Options of the kernel $\mathcal{K}(y)$ relevant to our problem are:
\begin{align}
\mathcal{K}(y) & = \frac{1}{(2 \pi)^{d/2}} \exp{\left( -\frac{1}{2} y^T y\right)} \quad & \text{Gaussian} \label{eqKernelGaussian} \\
\mathcal{K}(y) & = \frac{1}{2} c_d^{-1} (d+2)(1-y^T y) \quad & \text{Epanechnikov} \label{eqKernelEpanechnikov} \\
\mathcal{K}(y) & = \tfrac{1}{2}, \quad \text{for } |y| < 1 \quad & \text{Tophat}  \label{eqKernelTophat} 
\end{align}
where $c_d$ is the volume of a unitary sphere in $d$ dimensions ($c_1 = 2$, $c_2 = 2 \pi$, $c_3 = 4 \pi /3$). The choice of kernel ultimately has a minor effect on accuracy, which depends much more strongly on the bandwidth. The bandwidth required for kernel density estimation was computed based on Silverman's bandwidth selection factor $A(\mathcal{K})$ as
\begin{align}
A(\mathcal{K}) & = \left( \frac{4}{2d+1} \right)^{\frac{1}{d+4}}  \quad & \text{Gaussian} \\
A(\mathcal{K}) & = \left( \frac{8d(d+2)(d+4)(2 \sqrt{\pi})^d}{(2d+1)c_d} \right)^{\frac{1}{d+4}}  \quad & \text{Epanechnikov} 
\end{align}
The choice of the kernel should also be made considering such traits as differentiability and computational efficiency \cite{silverman_density_1986}; those aspects will be discussed in detail in Sec. \ref{sec:tests}. 

\subsection{Consistency criteria} 

\label{Sec:consistency}

The transfer of information from the continuum plasma code to the discrete material code must be done in such a way that the conservation laws are satisfied at the interface between the plasma and the material. The interface itself does not have to artificially produce mass, momentum, and energy, or in other words, the numerical method employed for the conversion from continuum to discrete (both forward and backward) does not have to produce artificial moments of the distribution functions up to a given order $k$. Considering the velocity moments up to third order at the surface,
\begin{align}
& n_s = \int f_s d\textbf{v}, \quad \Gamma_s = \int \textbf{v} f_s  d\textbf{v},\\ 
& \textbf{P}_s = \int m_s \textbf{v} \textbf{v} f_s  d\textbf{v}, \quad \textbf{Q}_s = \int \frac{1}{2} m_s v^2 \textbf{v} f_s d\textbf{v}, \quad \text{at } G(\textbf{x}) = 0
\end{align}
consistency is ensured when the artificial moments generated by the conversion of the true distributions $f_s$ into their density estimates $\hat{f}_s$ tend toward zero:
\begin{align}
\Delta n_s & = \int (\hat{f}_s - f_s) d\textbf{v} \rightarrow 0, \quad \text{at } G(\textbf{x}) = 0  \label{EqConsistency1}  \\ 
\Delta \Gamma_s & = \int \textbf{v} (\hat{f}_s - f_s)  d\textbf{v} \rightarrow 0, \quad \text{at } G(\textbf{x}) = 0 \\
\Delta \textbf{P}_s & = \int m_s \textbf{v} \textbf{v} (\hat{f}_s - f_s)  d\textbf{v}  \rightarrow 0, \quad \text{at } G(\textbf{x}) = 0 \\
\Delta \textbf{Q}_s & = \int \frac{1}{2} m_s v^2 \textbf{v} (\hat{f}_s - f_s) d\textbf{v} \rightarrow 0, \quad \text{at } G(\textbf{x}) = 0 \label{EqConsistency2}
\end{align}
In order to enforce such consistency at the interface, the common factor $\hat{f}_s - f_s$ appearing in Eqs.~\ref{EqConsistency1}-\ref{EqConsistency2} must be numerically convergent. We considered the Mean Integrated Squared Error ($MISE$) as a measure of optimal density estimate for such convergence,
\begin{equation}
MISE(\hat{f}) = \mathop{\mathbb{E}} \int (\hat{f}(\vect{x}) - f(\vect{x}))^2 d \vect{x}
\label{eq:mise}
\end{equation}
$MISE$ is simply the expected value of the integrated squared error. As the $MISE$ measure is not a dimension-free quantity, and as such it is not suitable to compare results of different dimensionality, we constructed  a dimensionless quantity 
\begin{equation}
ISE(\hat{f}) = \frac{ \int (\hat{f}(\vect{x}) - f(\vect{x}))^2 d\vect{x} }{ \int (f(\vect{x}))^2 d\vect{x} }
\label{eq:miseNormalized}
\end{equation}
by introducing the scaling factor $\int f^2 d\vect{x}$ as originally suggested by Epanechnikov \cite{Epanechnikov1969}. However, the error measure of Eqs.~\ref{eq:mise}-\ref{eq:miseNormalized} can be used only when the true distribution $f(x)$ is known, such as in analytical verifications, making the ISE impossible to calculate for all cases where analytical solutions are not available. More generally for our case, a purely data-based estimation of $MISE$ and $ISE$ is desired, and may be calculated via bootstrapping \cite{efron_jackknife_1980}. 
In the present work, the following procedure has been used for the calculation of the bootstrapped $MISE$ (Faraway and Jhun \cite{faraway_bootstrap_1990}): 
\begin{enumerate}
\item Obtain the $N_{ex}$ samples of particles leaving the surface, $\vect{q}_1, \vect{q}_2, ... , \vect{q}_{N_{ex}}$ 
\item Estimate the density function of the samples, $\hat{f}$, using one of the kernels $\mathcal{K}(y)$
\item Draw $b$ bootstrap samples $(\vect{q}^*_1, \vect{q}^*_2, ... \vect{q}^*_b)$ from $(\vect{q}_1, \vect{q}_2, ... \vect{q}_{N_{ex}})$ with replacement
\item Generate $b$ bootstrap estimates $\hat{f}^*_b$ from $(\vect{q}^*_1, \vect{q}^*_2, ... \vect{q}^*_b)$
\item Define the bootstrapped $ISE$, denoted $\widehat{ISE}$, as: 
\begin{equation}
\widehat{ISE} = \frac{ \sum^b_{l=1} \int (\hat{f}^*_l(\vect{x}) - \hat{f}(\vect{x}))^2 d\vect{x}}{b \int \hat{f}(\vect{x})^2 d \vect{x} }
\end{equation}
where the integral in the denominator is the normalization factor.
\end{enumerate}
If the density estimates used to construct $\hat{f}$ and $\hat{f}^*_B$ are convergent, the estimate $\widehat{ISE}$ tends toward the value of the true $ISE$, that is $\widehat{ISE} \rightarrow ISE$, and an estimate of the error may therefore be calculated without any knowledge of the true distribution from which the original samples were drawn. The convergence properties of the $\widehat{ISE}$ have been numerically characterized in the next section as a function of the sample size of discrete particles. The computational time for each numerical test is also reported for a direct comparison of the computational efficiency of the Density Estimation techniques adopted in this work.

\section{Numerical Tests}
\label{sec:tests}

\subsection{Convergence tests and timing}

For the dynamically coupled model intended for this work, density estimation must be performed frequently and on-the-fly, in the most demanding case at each time step of the Boltzmann solver. The possibility of constructing distributions on-the-fly by using particle data strictly depends on the convergence rate and on the computational cost of the density estimation algorithm; a better convergence rate means that a smaller number of Markov chains is necessary to reconstruct the distributions. Since this issue is of significant practical importance, it will be addressed and discussed here in detail. In this section we report the numerical tests aimed at characterizing the convergence behavior and the timing of the coupling scheme of Sec.~\ref{sec:equations}. 

Two types of tests were performed: (1) first, analytical tests of verification using samples drawn from a known Probability Distribution Function (PDF), so that the mean integrated square error ($MISE$) of each estimate could be calculated analytically; (2) second, the estimation techniques were applied to distributions of sputtered particles obtained from the surface erosion model of Sec.~\ref{sec:bca}. Since in this second case the distributions of sputtered particles have no analytic solution, the $\widehat{ISE}$ was estimated through the bootstrapping algorithm described in Sec.~\ref{Sec:consistency}.

\subsubsection{Analytical tests}

\begin{figure}
	\centering
	\includegraphics[width=0.85\textwidth]{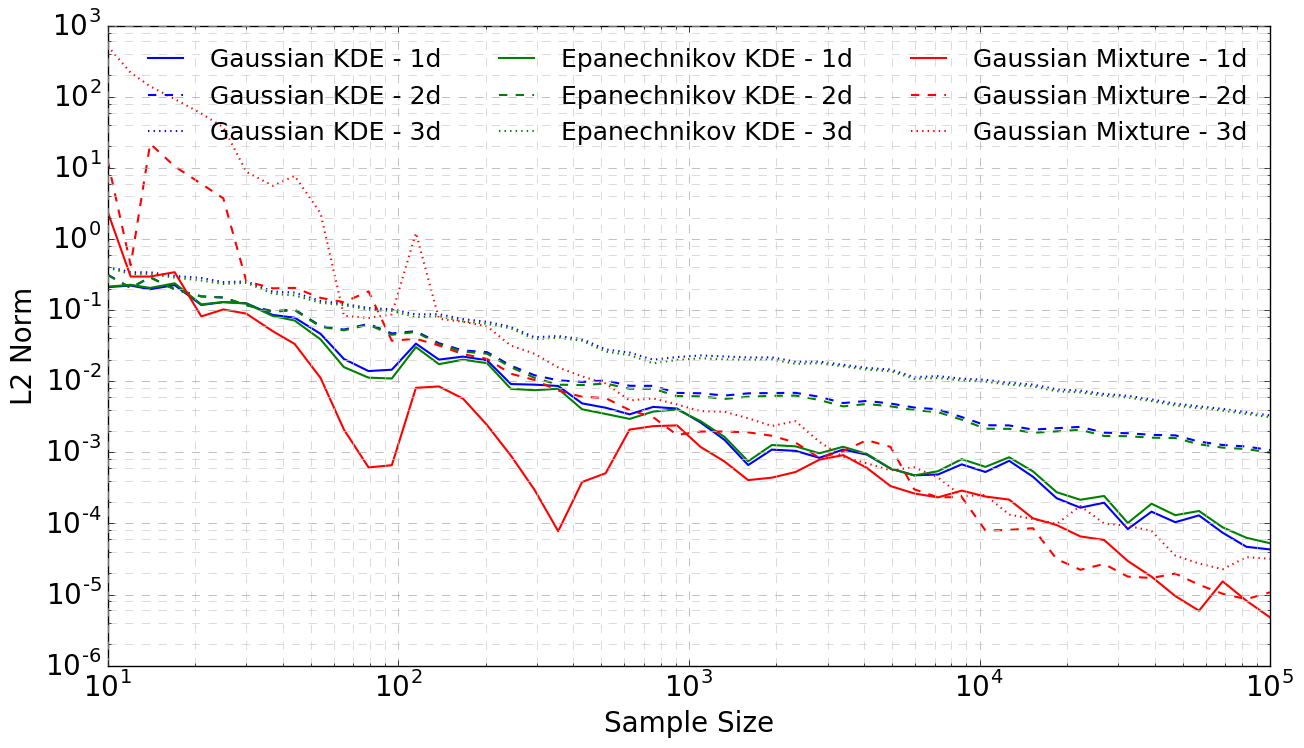}
	\caption{Integrated squared error for the density estimates as a function of sample size. Samples were drawn and estimates were fitted in one through three dimensions. The solid lines are the one-dimensional calculations, the dashed lines are two-dimensional, and the dotted lines are three-dimensional. }
	\label{fig:ise_comp}
\end{figure}

\begin{figure}
	\centering
	\includegraphics[width=0.85\textwidth]{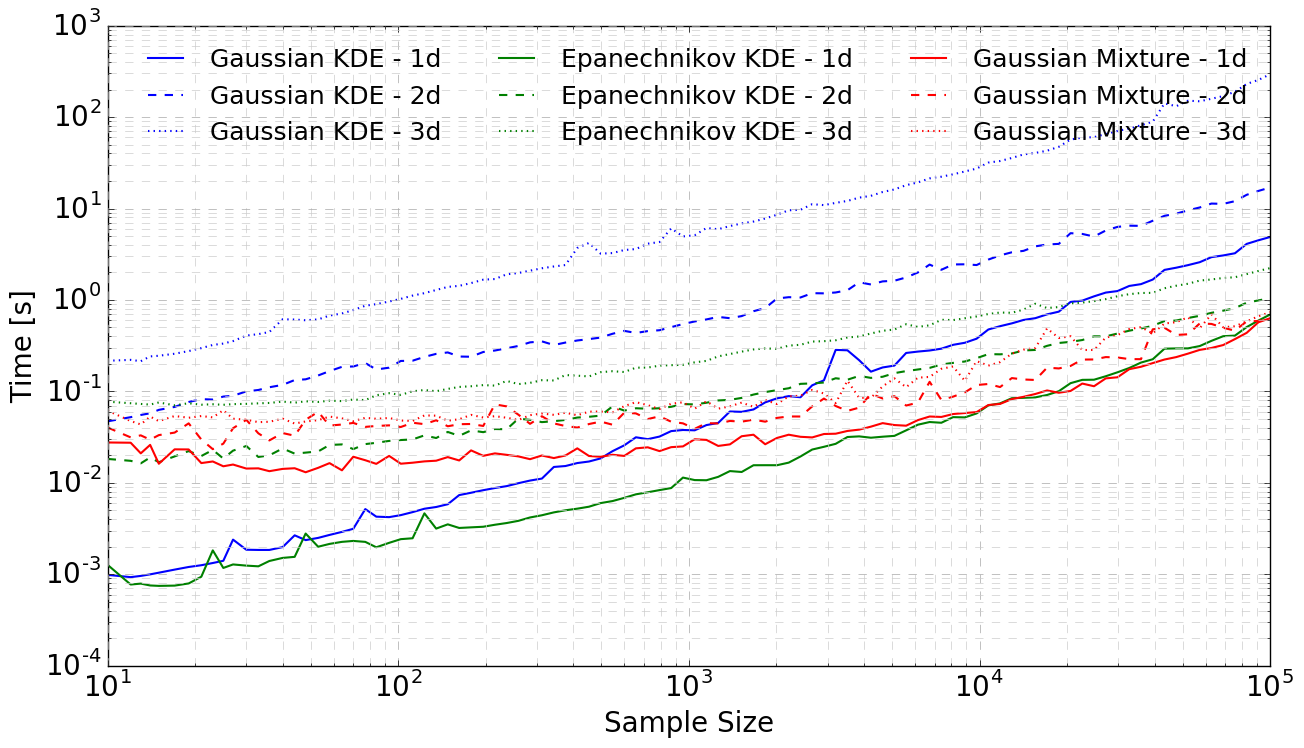}
	\caption{Computational time for the density estimates as a function of sample size. Samples were drawn from the standard Gaussian distribution in one (a), two (b), and three (b) dimensions.}
	\label{fig:gaussian_time}
\end{figure}

The analytical tests were performed using samples drawn from 1D, 2D, 3D unbiased unimodal Gaussian PDF distributions, which were reconstructed on a uniform grid of $N^d = 60^d$ gridpoints ($d=1,2,3$ dimensions) using both the GMM method (Eq.~\ref{eq:gmm}) and the KDE methods (Eqs.~\ref{eq_kde}--\ref{eqKernelEpanechnikov}, Gaussian-KDE and Epanechnikov-KDE). The results are reported in Figures \ref{fig:ise_comp} and \ref{fig:gaussian_time} showing the convergence rate (Fig.~\ref{fig:ise_comp}) and the computational time (Fig.~\ref{fig:gaussian_time}) as a function of the particle number $N_{mc}$. The most evident trend in the convergence rate plot (Fig.~\ref{fig:ise_comp}) is the ``curse of dimensionality'' of a probability distribution of dimension $d$. As expected \cite{nagler_evading_2016}, increasing the dimensionality of the problem both increases the amount of error in the system and decreases the rate of convergence. As can be seen from the same figure, increasing the sample size from $10^1$ to $10^5$ in the one-dimensional Gaussian-KDE case decreases the error from 20\% to 0.008\%; in three dimensions, the same density estimate only decreases the $\widehat{MISE}$ from 30\% to 0.3\%. The GMM model has a less drastic decrease in convergence rate. However, while noticeably better with large sample sizes, the GMM performs worse in all three dimensions at a low sample size than the KDE methods; it only performs better than the KDE when the sample sizes are greater than $\sim$15, 30, and 100 in one, two, and three dimensions, respectively. The computational time for the density estimates (Fig.~\ref{fig:gaussian_time}) increases with both the number of particles $N_{mc}$ and the dimensionality of the problem, with approximately an order of magnitude increase per dimension. Notably, the cost of the GMM model only marginally increases with dimensionality. At the largest sample size of $10^5$, the Gaussian-KDE is two orders of magnitude slower than both the Epanechnikov-KDE and the GMM, mainly due to the presence of the exponential factor, causing each call to the Gaussian-KDE function to be more computationally expensive than the Epanechnikov-KDE function. 
Additional tests were performed to characterize the convergence of the bootstrapped $\widehat{ISE}$ to the analytical $ISE$ as a function of the number of samples. 
The tests revealed deviations smaller than 1\% for a sample size $N_{mc} >30$. With 50 samples, the accuracy falls below 0.5\%. Similar tests revealed that approximatively 50 bootstrapped samples were sufficient for an estimate of the distribution of sputtered particles, as expected from a ``well-behaved'' smooth and unimodal underlying distribution.

\subsubsection{Distributions of sputtered particles}

\begin{figure}
	\centering
	\includegraphics[width=0.85\textwidth]{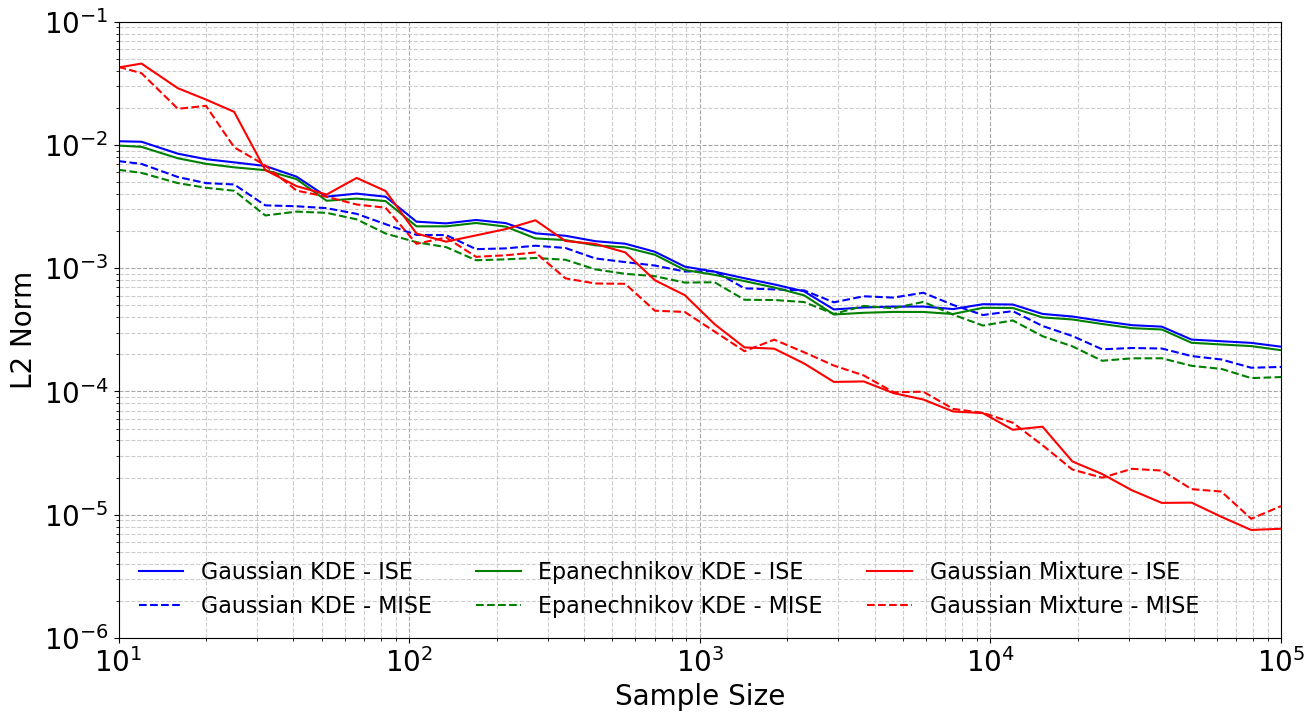}
	\caption{Integrated Squared Error compared to the bootstrapped Mean Integrated Squared Error ($\widehat{MISE}$) in three dimensions. The dotted line is the $\widehat{MISE}$ estimate, and the solid lines are the $ISE$ shown in Fig.~\ref{fig:ise_comp}.  }
	\label{fig:3d_ise_mise_comp}
\end{figure}

\begin{figure}
	\centering
	\includegraphics[width=0.85\textwidth]{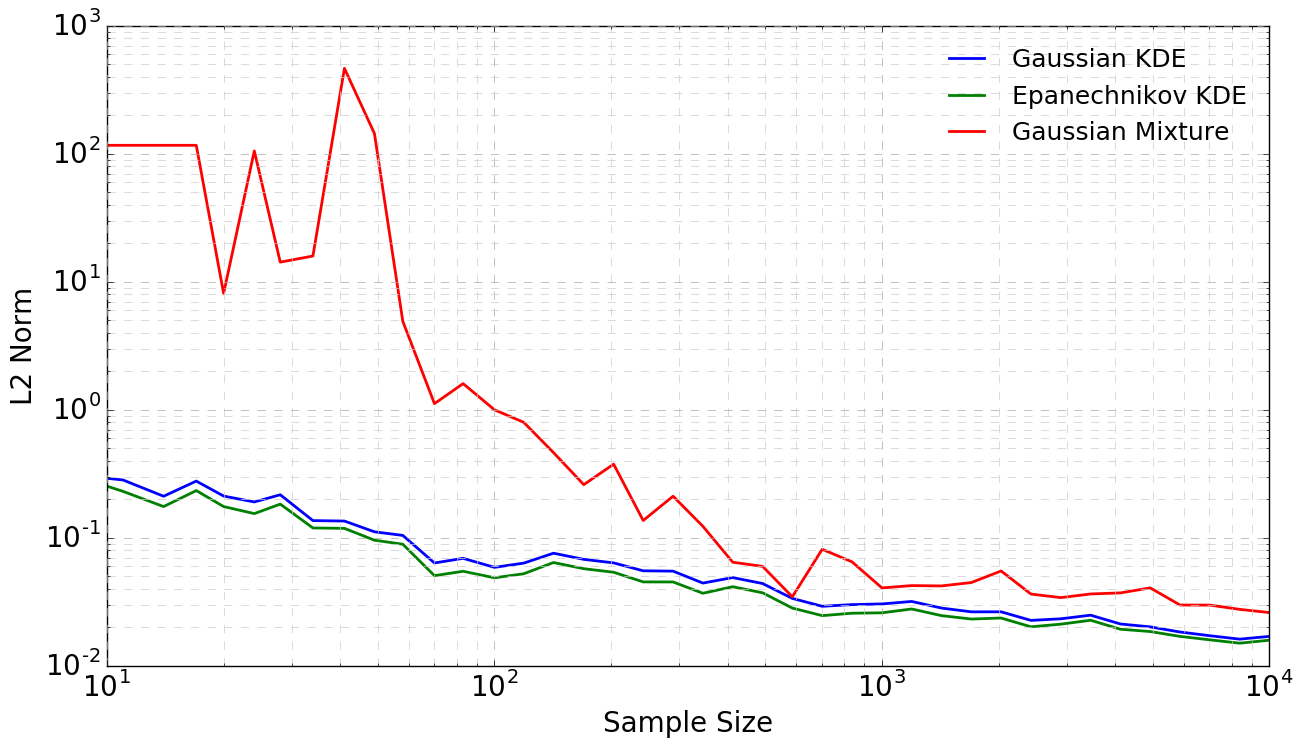}
	\caption{L2-norm of the Integrated Square Error ($\widehat{ISE}$) of the continuum distributions reconstructed from discrete data of sputtered particles, calculated using three different Density Estimation techniques. The $x$ axis reports the number of sputtered particles, ranging from 10 to $10^4$. }
	\label{fig:3d_mise_tridyn}
\end{figure}

\begin{figure}
	\centering
	\includegraphics[width=0.85\textwidth]{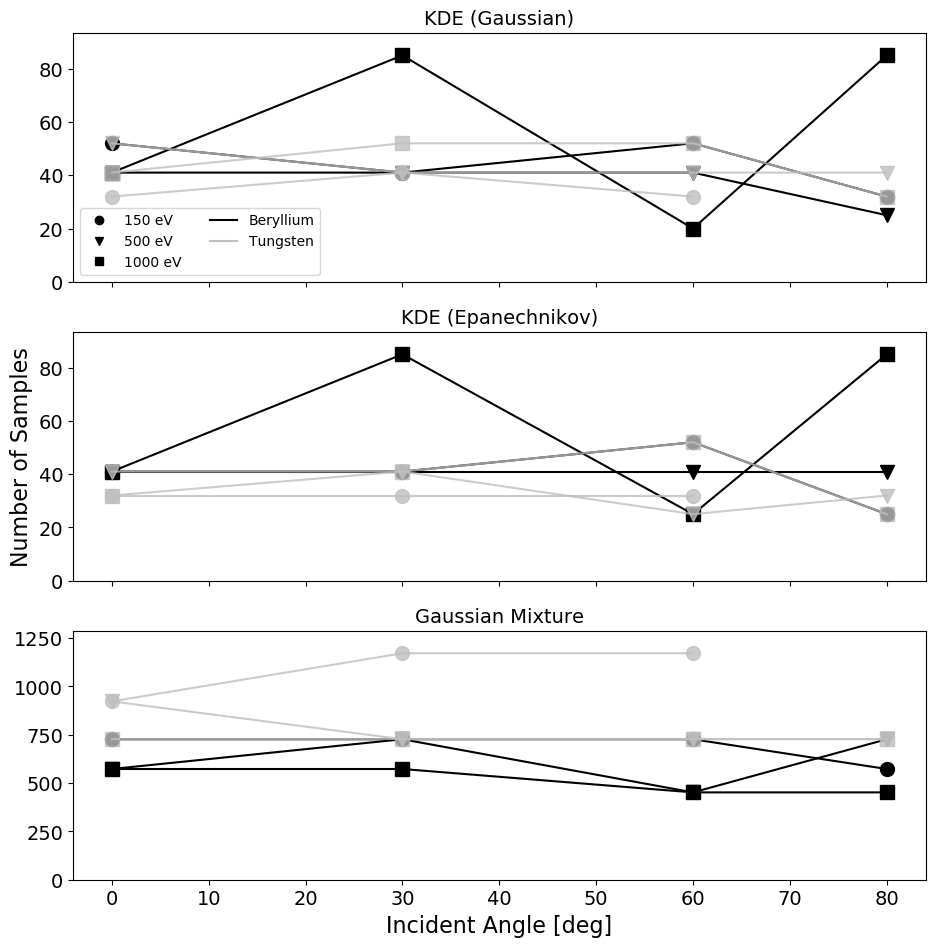}
	\caption{Comparison of the number of samples required to obtain an $\widehat{ISE}$ of 10\% for estimated beryllium and tungsten sputtered distributions. Incident angles between 0-80 degrees and energies of 150 eV, 500 eV, and 1000 eV were considered for both materials.}
	\label{fig:error_comp}
\end{figure}

Tests on distributions of sputtered particles (Fig.~\ref{fig:3d_ise_mise_comp}) were performed using data samples from the surface erosion model of Sec.~\ref{sec:bca}, calculated with the F-TRIDYN code \cite{drobny_f-tridyn:_2017}. 
The sputtering simulations were run over a range of parameters of typical interest in nuclear fusion, considering a helium plasma incident on a tungsten surface at energies comprised between $\epsilon = 10 - 1000$ eV and angles of incidence between $\theta = 0^{\circ}-90^{\circ}$ (ranging between normal incidence and grazing incidence).  
Sample sizes ranging from $N_{mc} = 10^1$ to $10^4$ were drawn from the particle data for the calculation of the bootstrapped $\widehat{ISE}$. Density estimates were performed also in this case using both the GMM (Eq.~\ref{eq:gmm}) and the KDE  (Eqs.~\ref{eq_kde}--\ref{eqKernelEpanechnikov}) methods. The distributions were reconstructed on a three-dimensional velocity grid of $N^{d=3} = 60^3$ grid points. Figure \ref{fig:3d_mise_tridyn} (left) shows the results of the numerical tests for the case of He impacting on W at energy $\epsilon = 1000$ eV and angle $\theta = 60^{\circ}$). The figure reports the bootstrapped $\widehat{ISE}$ for the three density estimation techniques under analysis. The tests show that the mean integrated square error of the GMM with small sample sizes is much larger than the KDE methods. The larger error of the GMM case can be attributed to the presence of both a tail in the data and skewness of the distribution along the direction perpendicular to the surface, both of which represent significant deviation from Gaussian distributions. Both KDE methods perform better in this case since they do not rely on a parametric fit and can resolve these features with a smaller particle sample relative to the GMM. The plot also shows the convergence properties of the different techniques. 
Figure \ref{fig:error_comp} summarizes a large set of simulations at different energies and angles for two wall materials, beryllium and tungsten. The figure reports the number of samples required to obtain an $\widehat{ISE}$ of 10\% for the estimation of beryllium and tungsten sputtered distributions. Incident angles between 0-80 degrees and energies of 150 eV, 500 eV, and 1000 eV were considered for both materials. While GMM techniques require 500-1000 particle samples, the two KDE techniques have superior convergence, allowing to obtain an integrated $\widehat{ISE}$ error of the order of 10\% using a small sample of 40-80 sputtered particles. For such small particle samples of the order of $10^2$, the computational time required for density estimation largely depends on the type of technique adopted (see  Fig.~\ref{fig:gaussian_time}), with the Epanechnikov-KDE being a best choice for the reconstruction of 3D distributions. 

\subsubsection{Tests outcome}

The numerical tests revealed the superior convergence properties of Kernel Density Estimation methods over  Gaussian Mixture Models. KDE methods allow the reconstruction of the distributions with a smaller number of particle samples. Three dimensional distributions of sputtered particles could be reconstructed with a bootstrapped integrated square error $\widehat{ISE}$ smaller than 0.5\% using a sample of just 50 particles. However, the two KDE methods differ considerably in computational time, with the Epanechnikov-KDE being up to two orders of magnitude faster than the Gaussian-KDE. Such a difference is less dramatic in the range of interest of small sample sizes (approximately a factor of 2), but still showing that the Epanechnikov Kernel Density Estimator is the most appropriate method for on-the-fly reconstruction of distributions of sputtered particles.  

\subsection{Sheath formation in front of a wall releasing impurities}

The methodology presented in the previous sections has been applied to  the study of microscopic erosion of material walls exposed to a high-density magnetized plasmas. 
Here we report a numerical example, showing a 1D3V Boltzmann simulation of plasma sheath formation and depletion in front of a grounded wall in a strongly-magnetized environment. A magnetic field $B$  almost parallel to the wall was chosen, with magnetic angle $\psi = 86^{\circ}$ with respect to the normal to the surface. The simulation was for a helium plasma on a beryllium wall, and it was including four Boltzmann species ($s=4$): He$^+$ (He ions), He-I (He neutrals), Be$^+$ (Be ions), Be-I (Be neutrals). The plasma is allowed to propagate within a one-dimensional domain of size L = 1 mm in physical space and within $\pm 5 v_{th}$ in the velocity space, where $v_{th}$ is the thermal velocity of each species, $v_{th} = (k_b T_s/M_s)^{\frac{1}{2}}$, with $T_s$ and $M_s$ are the temperature and mass of each species respectively. The plasma was initialized with Maxwellian distributions for He$^+$ and He-I, and no Be species (Be$^+$ and Be-I) were initially present. 

\begin{figure}[h]
\subfloat[He Ions] {
 \includegraphics[width=0.38\textwidth]{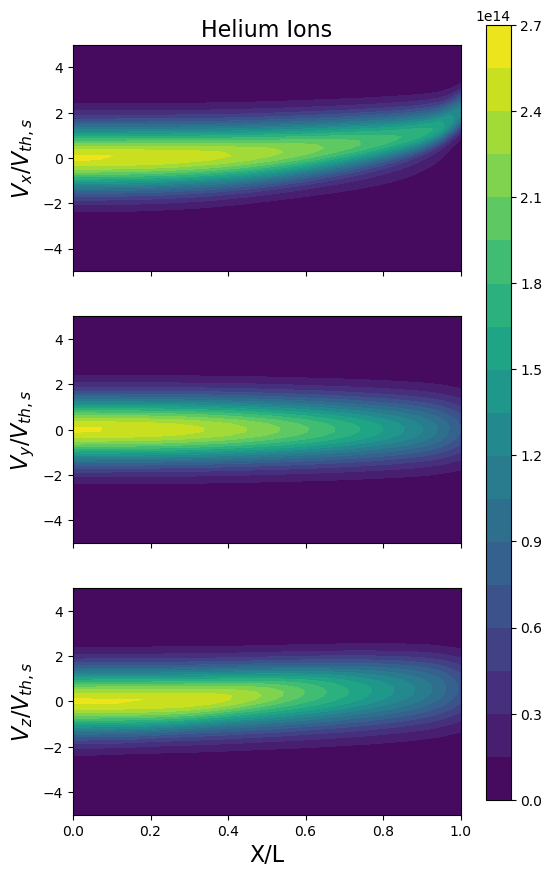}
 \label{fig:he_ions_3v}
}
\hfill
\subfloat[He Neutrals]{
  \includegraphics[width=0.38\textwidth]{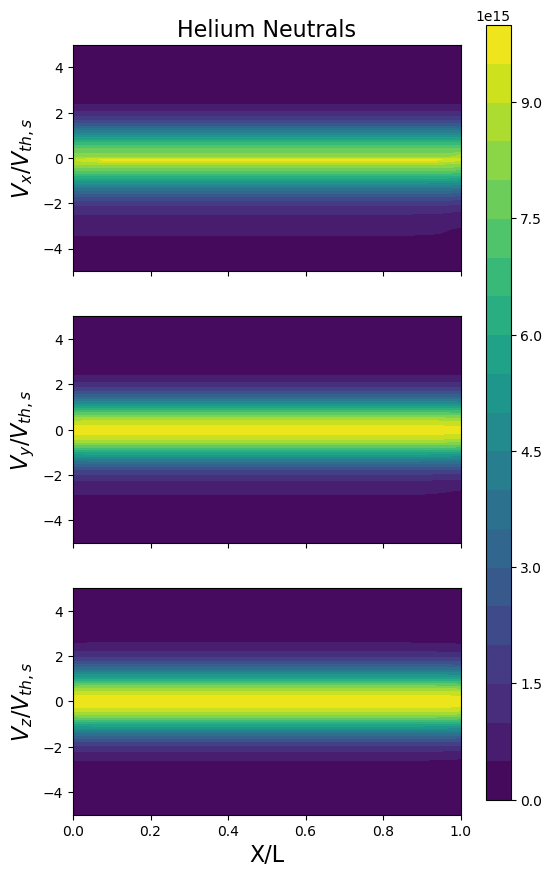}
  \label{fig:he_neutrals_3v}
}

\subfloat[Be Ions]{
	\includegraphics[width=0.38\textwidth]{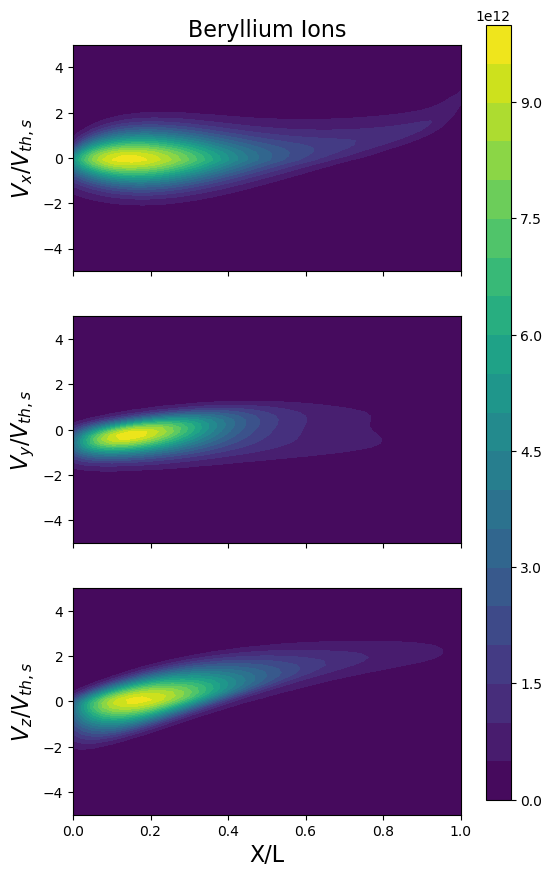}
	\label{fig:be_ions_3v}
}
\hfill
\subfloat[Be Neutrals]{
	\includegraphics[width=0.38\textwidth]{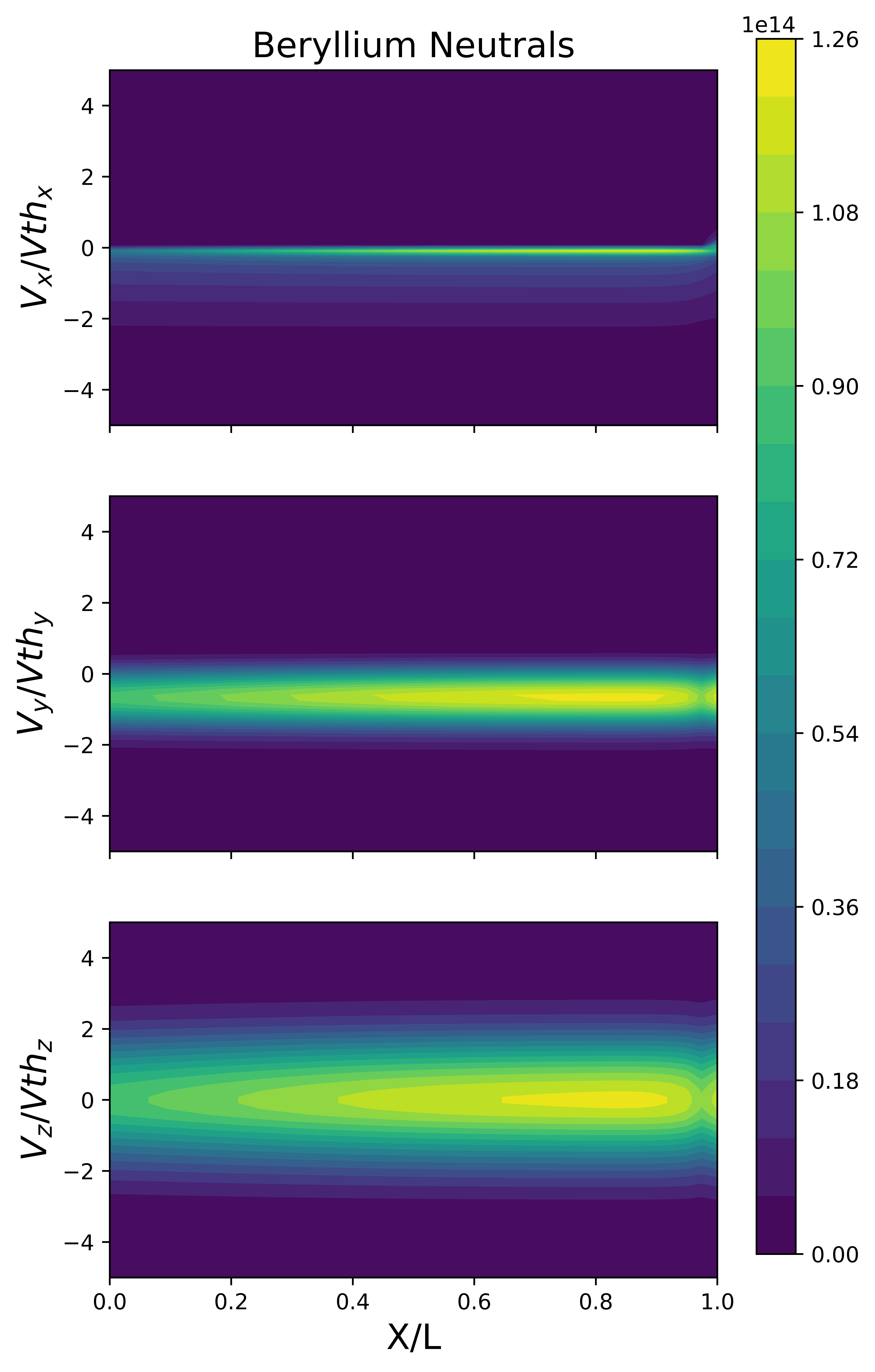}
	\label{fig:be_neutrals_3v}
}

\caption{Phase space of He ions (a), He neutrals (b), Be ions (c), and Be neutrals (d) in a 1D3V magnetized plasma after 30 ns. In each, the top figure is the $X-V_x$ plane, the middle is the $X-V_y$ plane, and the last is the $X-V_z$ plane. \label{fig:1d3v_phasespace_1}}

\end{figure}

Figure \ref{fig:1d3v_phasespace_1} shows a snapshot of the four phase spaces of each species during the plasma sheath formation. Several striking features have been observed. The helium ions (Fig.~\ref{fig:1d3v_phasespace_1}.a) are supersonically accelerated toward the surface, and are responsible for the release of beryllium impurities in neutral state (Fig.~\ref{fig:1d3v_phasespace_1}.d). The Be neutrals are able to cross the Debye sheath and the magnetic presheath mostly remaining in neutral state. They ionize more frequently as they pass from the sheath region to the plasma bulk, turning into beryllium ions (Fig.~\ref{fig:1d3v_phasespace_1}.c). Once ionized, the beryllium impurities become electrically and magnetically driven by the plasma; a fraction of the Be+ population flows back toward the wall (redeposition) and the remaining fraction flows toward the plasma bulk contributing to plasma contamination. Both Be$^+$ and neutrals are symmetric in the $V_{thy}$ and $V_{thz}$ dimensions, although a slight acceleration in $v_z$ of the Be ions due to the magnetic field is visible. One of the most striking physics features is related to the effect of magnetization. The magnetization of the plasma causes an $E \times B$ acceleration along the $v_z$ direction, which is in turn responsible of the modification of the energy-angle distribution functions at the wall. Finally, the velocity of the particles incident on the walls is smaller than the classical value of $2.3 V_{th}$, suggesting that the electrostatic portion of the sheath (Debye sheath) might decrease in size as the magnetic angle increases.

\section{Conclusions}

In this work we have numerically characterized two classes of density estimation techniques useful for multiscale coupling of kinetic plasma-material interaction models, namely the Gaussian Mixture Model technique and Kernel Density Estimation techniques. First, we have constructed a fully-kinetic model of the plasma material interface, including: (1) a continuum 1D3V plasma solver of the multi-species Vlasov-Poisson-BGK problem, and (2) a discrete ion-matter-interaction model based on Binary Collision Approximation. Then, we have constructed a coupling technique based on Density Estimation, and defined a consistent strategy to quantify the error introduced by such coupling, based on bootstrapping of a normalized Mean Integrated Square Error.

From the numerical tests we have found that Kernel Density Estimation (KDE) techniques have superior convergence properties compared to Gaussian Mixture models, and are thus preferable for the treatment of distribution functions exhibiting significant deviations from a Maxwellian, such as those encountered in sputtering problems. Among the KDE methods, we have observed that Epanechnikov's KDE method offers a good compromise between convergence rate and computational time, and it is thus the method of preference for distribution reconstruction. Finally, we have applied the kinetic model to a case of interest for fusion applications, simulating a multi-species strongly-magnetized plasma facing a material wall releasing impurities.

\section{Acknowledgments}

This work was funded by the US Department of Energy through Grant No. DE-SC0004736. Numerical tests utilizing density estimation techniques were performed in Python 2.7 with the \texttt{scikit-learn} library \cite{pedregosa_scikit-learn:_2011}.



\newpage
\clearpage

\end{document}